\documentclass[twocolumn,showpacs,preprintnumbers,amsmath,amssymb,prl]{revtex4}

\usepackage{amssymb}
\usepackage{graphicx}
\usepackage{bm}

\begin{document}
\title{Evolution of critical scaling behavior near a ferromagnetic quantum phase transition}

\author{N. P. Butch}
\altaffiliation{Present address: Center for Nanophysics and Advanced Materials, University of Maryland, College Park, MD 20742}
\email{nbutch@umd.edu}
\author{M. B. Maple}
\affiliation{Department of Physics and Institute for Pure and Applied Physical Sciences, University of California, San Diego, La Jolla, CA 92093}
\date{\today}

\begin{abstract}
Magnetic critical scaling in URu$_{2-x}$Re$_{x}$Si$_{2}$ single crystals continuously evolves as the ferromagnetic critical temperature is tuned towards zero via chemical substitution.  As the quantum phase transition is approached, the critical exponents $\gamma$ and $(\delta-1)$ decrease to zero in tandem with the critical temperature and ordered moment, while the exponent $\beta$ remains constant.  This novel trend distinguishes URu$_{2-x}$Re$_{x}$Si$_{2}$ from stoichiometric quantum critical ferromagnets and appears to reflect an underlying competition between Kondo and ferromagnetic interactions.
\end{abstract}

\maketitle

Understanding the quantum critical phenomena found in the vicinity of zero-temperature phase transitions \cite{Sachdev00,Coleman05} remains a principal challenge in the study of correlated electron materials such as high-temperature superconductors and heavy fermion compounds. In the vicinity of these quantum phase transitions (QPTs), quantum mechanical fluctuations, as opposed to thermal fluctuations, can give rise to dramatic novel effects, including non-Fermi liquid behavior, that are observed experimentally in numerous physical systems \cite{Maple95,Lohneysen07}. In contrast to these well-documented phenomena, there are few experimental characterizations of the critical scaling of the order parameter itself at QPTs. We report a study of magnetic critical scaling in single crystals of the heavy fermion ferromagnet URu$_{2-x}$Re$_{x}$Si$_{2}$, which exhibits an unprecedented trend as the magnetic ordering temperature is tuned towards zero via chemical substitution.

Recently, the properties of ferromagnetic (FM) heavy fermion compounds at finite temperature and at QPTs  have been the subject of much research. Among the best-known examples are ZrZn$_2$ and UGe$_2$, and for both, the Curie temperature $T_C$ can be suppressed by applied pressure, disappearing discontinuously at a first-order transition \cite{Uhlarz04,Pfleiderer02}. There are further indications, both theoretical \cite{Chubukov04} and experimental \cite{Uemura07} that FM QPTs are generally discontinuous, though this matter is not yet settled.  Heavy fermion ferromagnets are particularly complex, due to the coexistence of FM and Kondo interactions, and uncertainty whether light or heavy quasiparticles participate in the FM order.

Although the heavy fermion compound URu$_{2}$Si$_{2}$ has achieved notoriety for its unconventional superconductivity and mysterious hidden order phase, it is less known that the replacement of Ru with either Tc  or Re yields heavy fermion ferromagnetism \cite{Dalichaouch89,Dalichaouch90,Torikachvili92,Kohori93}. The itinerant FM order is characterized by a small moment and lacks discernable anomalies in electrical resistivity or specific heat at $T_{C}$.  In polycrystalline samples of URu$_{2-x}$Re$_{x}$Si$_{2}$, the maximum $T_{C} = 38$~K occurs at $x=0.80$ \cite{Dalichaouch89}, and $T_{C}$ decreases to 0~K for  $x<0.35$, although the FM phase boundary in this range of concentrations has not yet been accurately determined. Astoundingly, far in the FM ordered phase, non-Fermi liquid (NFL) behavior is observed in the low-temperature bulk properties \cite{Bauer05}. A recent neutron scattering study identified energy-temperature scaling in the dynamic magnetic susceptibility, which seems to be associated with the bulk NFL behavior \cite{Krishnamurthy08}. In light of these provocative findings, it is crucial to properly define the boundary of the FM phase and to confirm that the novel behavior observed in polycrystalline samples does not arise from orientational averaging of grains or the presence of defects.

In this study, we measured the bulk magnetization of single crystal samples of URu$_{2-x}$Re$_{x}$Si$_{2}$.  Systematic measurements on $0.20 \leq x \leq 0.60$ (10-30\% Re) and a subsequent magnetic scaling analysis lead to several exceptional results.  First, the magnetic critical behavior, in which critical exponents decrease continuously towards zero along with  $T_{C}$, is dissimilar to that of any known materials and unexplained by currently available theoretical models. Second, the previously identified NFL behavior is intimately related to the novel magnetic critical behavior. Finally, the FM phase boundary persists to lower $x$ than previously believed, in close proximity to the hidden order phase, suggesting that these two phases may even meet at a common critical point.

Single crystals of URu$_{2-x}$Re$_{x}$Si$_{2}$ were synthesized via the Czochralski technique in a tri-arc furnace and magnetization measurements were performed in a Quantum Design MPMS. The moderate magnetic anisotropy observed in URu$_{2}$Si$_{2}$ increases substantially upon Re substitution, ranging from about 10 to over 100 at higher concentrations. Consequently, all measurements were performed with the magnetic field applied along the [001] easy axis. Demagnetization corrections were less than 5\%.

\begin{figure}
    {\includegraphics[width=3.4in]{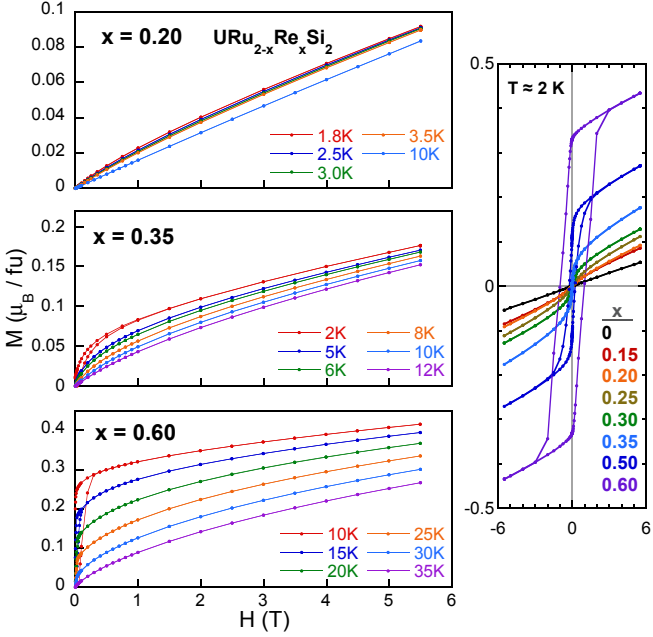}}
    \caption{Magnetization isotherms for several URu$_{2-x}$Re$_{x}$Si$_{2}$ samples exhibiting ferromagnetism. As Re concentration increases, the $M(H)$ curves reflect an increased $T_{C}$ and moment, which is shown on the right in a direct comparison of low-temperature data across the entire range of $x$ investigated. Data shown for $x=0.3$ and $x=0.6$ were measured at 1.8~K and the other data were taken at 2.0~K. }
    \label{MH}
\end{figure}

Representative isotherms of the magnetization as a function of magnetic field $M(H)$ are shown in Figure~\ref{MH} for several different Re concentrations for which URu$_{2-x}$Re$_{x}$Si$_{2}$ orders ferromagnetically.  As the Re concentration increases from $x=0.20$ to $x=0.60$, signatures of strengthening ferromagnetism are apparent: the magnitude of the magnetization grows, the $M(H)$ data exhibit more curvature, and magnetic hysteresis develops.  A direct comparison of low-temperature $M(H)$ curves emphasizes the emergence of magnetic order with increasing $x$, and the pronounced magnetic hysteresis at higher Re concentration.  The parent compound URu$_{2}$Si$_{2}$ exhibits linear paramagnetic $M(H)$ in this field range, but in URu$_{2-x}$Re$_{x}$Si$_{2}$, the isotherms gently curve and do not saturate.  Despite the clear increase in magnetization with Re concentration, even at $x=0.60$, the moment is a small fraction of the high-temperature paramagnetic effective moment of approximately 3.8~$\mu_\mathrm{B}$.

In the absence of anomalies in electrical resistivity and specific heat, the determination of $T_{C}$ at each $x$ is challenging, but is essential to the determination of the nominal QPT. The FM phase boundary of URu$_{2-x}$Re$_{x}$Si$_{2}$ was originally delineated in part by mean-field Arrott analysis \cite{Dalichaouch89}, although it was noted that a modified Arrott analysis produces unusual values of the scaling exponent $\delta$ \cite{Bauer05}. In this study, we have applied a more general FM critical scaling approach based on the Arrott-Noakes equation of state \cite{Arrott67}, which does not assume mean-field behavior. This procedure yields values of $T_{C}$, the ordered moment $M_0$ and the magnetic critical exponents $\beta$, $\gamma$, and $\delta$. The three exponents are typically defined for $M$, $H$, and reduced temperature $t = \frac{(T - T_{C})}{T_{C}}$ by $M \sim t^{\beta}$ for $t<0$, $M \sim H^{1/\delta}$ for $t=0$, $\frac{\partial M}{\partial H} = \chi \sim t^{-\gamma}$ for $t>0$, and $\delta-1 = \gamma/\beta$. Physically, $\beta$ describes how the ordered moment grows below $T_C$, with smaller values indicating faster growth; $\gamma$ describes the divergence of the magnetic susceptibility at $T_C$, with smaller values yielding sharper divergence; and $\delta$ describes the curvature of $M(H)$ at $T_C$, with smaller values reflecting less curvature and slower saturation.  The exponent $\delta$ and $T_C$ are first estimated by identifying the isotherm with constant power-law behavior over the widest range of $H$. The entire analysis requires the simultaneous satisfaction of three conditions: scaling in $|M|/|t|^{\beta}$ vs $H/|t|^{\delta\beta}$, i.e., the collapse of $M(H)$ data onto 2 diverging curves, for $t>0$ and $t<0$; modified Arrott plots of $|M|^{1/\beta}$ vs $|H/M|^{1/\gamma}$, i.e., linearized $M(H)$ isotherms evenly spaced in $t$; and, crucially, a consistent value of $T_C$ between scaling, modified Arrott, and power-law fitting methods.

\begin{figure}
    {\includegraphics[width=3.4in]{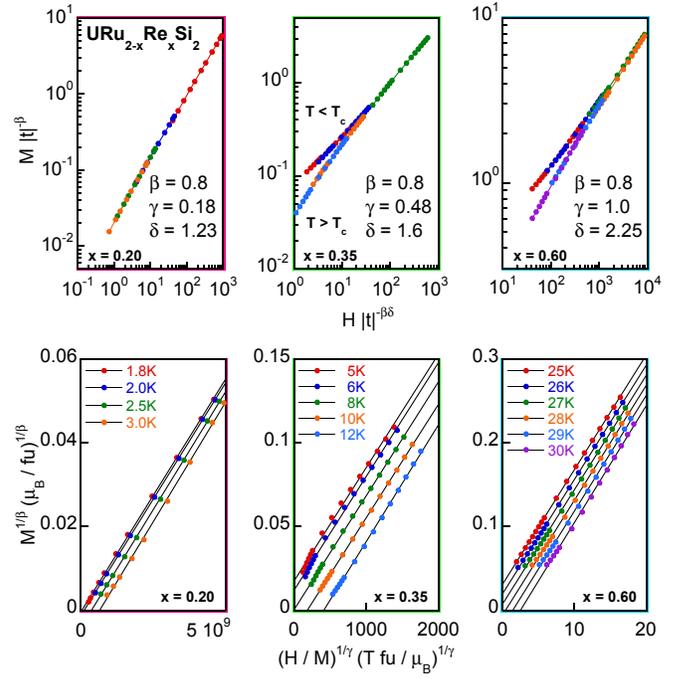}}
    \caption{Arrott-Noakes scaling law plots and modified Arrott plots, vertically paired, for several ferromagnetic samples of URu$_{2-x}$Re$_{x}$Si$_{2}$. In the scaling plots, $|M|/|t|^{\beta}$ vs $H/|t|^{\delta\beta}$, the $M(H)$ data collapse onto two curves for $T>T_C$ and $T<T_C$. In the modified Arrott plots, $|M|^{1/\beta}$ vs $|H/M|^{1/\gamma}$, the $M(H)$ data are linearized and evenly spaced in temperature. Each pair of plots yields a consistent set of values of critical exponents and $T_C$.}
    \label{scaling}
\end{figure}

Figure~\ref{scaling} exhibits the results of the scaling analysis for the $M(H)$ data shown in Fig.~\ref{MH}, which has been applied successfully to samples with $0.20 \leq x \leq 0.60$. Plotting scaled magnetization $|M|/|t|^{\beta}$ vs scaled field $H/|t|^{\delta\beta}$ collapses the $M(H)$ data within the critical $T$ range onto two curves, corresponding to $T > T_{C}$ and $T < T_{C}$. In the modified Arrott plots, the $M(H)$ isotherms are transformed into parallel lines evenly spaced in $T$ in a plot of $|M|^{1/\beta}$ vs $|H/M|^{1/\gamma}$, allowing a precise determination of the values of $T_C$.  The value of $M_0$ is estimated by extrapolating the $|M|^{1/\beta}$ intercepts to $T=0$. The value of $T_C$ for $x=0.20$ approaches 1.8~K, the lowest temperature accessible in this study. The consistency between the data and the Arrott-Noakes equation demonstrates the existence of FM order for $0.20 \leq x \leq 0.60$.

\begin{figure}
    {\includegraphics[width=3.4in]{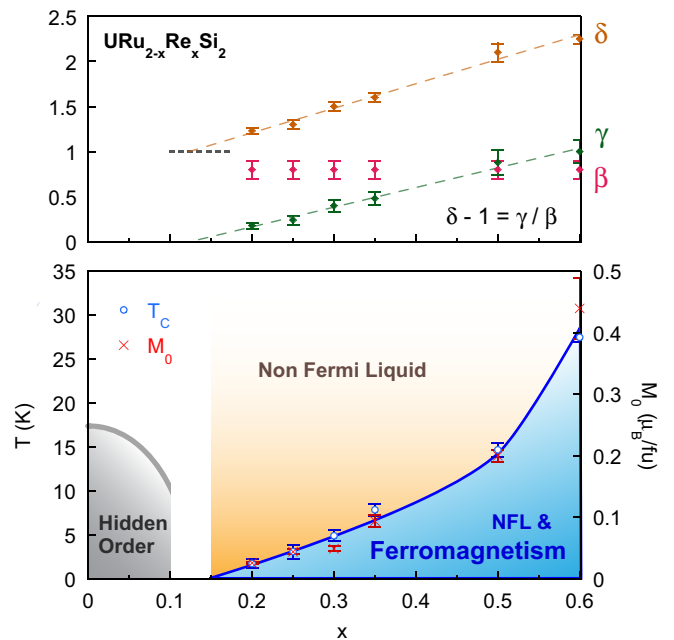}}
    \caption{Concentration dependence of magnetic critical exponents and phase diagram of  URu$_{2-x}$Re$_{x}$Si$_{2}$. Error bars denote the range of values that satisfy the scaling analysis. The exponents $\delta-1$ and $\gamma$, $T_C$, and $M_0$ all extrapolate to zero near $x=0.15$. The $x$-dependence of the exponents $\delta$ and $\gamma$ is well-described by a linear fit, which also holds true for $T_C$ and $M_0$ for $x \leq 0.5$; the FM phase boundary drawn is a guide to the eye.  The NFL regime, identified in measurements on polycrystals \cite{Bauer05,Krishnamurthy08}, extends into the FM phase. The hidden order phase boundary follows Ref. \cite{Bauer05}.}
    \label{phsdgm}
\end{figure}

The evolution of the critical exponents, $T_C$, and $M_0$ with Re concentration is presented in Fig.~\ref{phsdgm}. The $x$-dependence of the exponents is linear for all $x$, while a linear $x$-dependence of $T_C$ and $M_0$ persists to  $x=0.50$.  Whereas in most quantum critical systems only $T_C$ is tracked to zero, it is rather remarkable in URu$_{2-x}$Re$_{x}$Si$_{2}$ that $T_{C}$, $M_{0}$, $\gamma$ and $(\delta-1)$ all extrapolate to $0$ at $x = 0.15\pm0.03$. To our knowledge, this is the first example of such behavior. In particular, the continuous evolution of the exponents with $x$ seems to violate the tenet that one set of critical exponents controls the finite-temperature transitions, while possibly another set of exponents describes the quantum transition \cite{Hertz76}.  The presently determined critical exponents deviate substantially from mean-field values $(\beta = 0.5$, $\gamma = 1$, $\delta = 3)$ and those describing classical FM systems, where $\beta < 0.5$ and $\delta > 3$.  Moreover, even for $x>0.5$, where  $T_C > 10$~K, the exponents do not assume mean-field or classical values. Note that the curvature of the isotherms cannot be accounted for by subtracting a paramagnetic component, such as a large heavy-electron Pauli susceptibility or a Curie-law susceptibility due to magnetic impurities, as these corrections lead to inconsistencies between the scaling and modified Arrott analyses.

The critical exponents describe the unusual evolution of FM order in URu$_{2-x}$Re$_{x}$Si$_{2}$. As shown in Fig.~\ref{MH}, with decreasing $x$, the $M(H)$ isotherms exhibit decreasing curvature, as $\delta \rightarrow 1$.   Meanwhile, the value of $\beta$ remains constant, intimating that the $T$-dependence of the moment increase is independent of the magnitude of $T_C$ and $M_0$, as would typically be expected. However, $\beta=0.8$  is unusually large, suggesting a slower increase of the moment than in classical or mean-field ferromagnets. Because $\beta$ is constant,  $\delta \rightarrow 1$ constrains $\gamma \rightarrow 0$, which is quite peculiar. Typically, $\gamma \geq 1$, and physically, a suppression of $\gamma$ to 0 indicates increasingly sharper curvature in the divergent susceptibility at $T_{C}$, which can be interpreted as a trend towards a first-order transition at the QPT.  This notion conflicts with the expectation that the phase boundary should terminate at a higher-order critical endpoint.  However, we stress that at this ostensible first-order endpoint, both $T_C$ and $M_0$ tend to 0, and in this limit there would be no discontinuity in the order parameter. A similar suppression is expected in the critical exponent associated with the correlation length.

Generally, itinerant FM transitions appear to be described well by mean field or classical theories, but there are few available experimental assessments of their critical behavior near a QPT.  Mean-field scaling is reported for both chemically-tuned \cite{Yoshimura85,Sokolov06,Huy07} and pressure-tuned systems \cite{Akazawa04,Uhlarz04}. In contrast, significant deviation from classical critical exponents has been documented mostly in chemical substitution studies \cite{UbaidKassis08,Itoh08}, and rarely in stoichiometric ferromagnets \cite{Kida08}. Nonetheless, the continuous suppression of critical exponents is so far unique to URu$_{2-x}$Re$_{x}$Si$_{2}$.

One important consideration is the possible interplay between magnetism and the heavy fermion state inferred from the large electronic specific heat measured in the FM phase of URu$_{2-x}$Re$_{x}$Si$_{2}$. Based on the scaling analysis, a substantial heavy fermion paramagnetic contribution to $M(H)$ can be ruled out. This conclusion may imply that the itinerant FM moment must arise from a heavy quasiparticle band, although it is possible that the FM moment arises from a lighter band and that the large specific heat has an alternate source. In fact, both light and heavy band models of FM Kondo lattices predict a specific heat enhancement \cite{Perkins07,Yamamoto08}.  In either scenario, Re substitution regulates a delicate balance between Kondo and FM interactions.

The presence of fluctuations besides those of the order parameter are known to strongly influence critical scaling \cite{Ferreira05,Belitz05}. Disorder due to Re substitution may give rise to such fluctuations, but quantitative agreement with theoretical predictions for disordered itinerant ferromagnetism \cite{Belitz01} remains tentative. However, the now-apparent proximity of the FM phase to the hidden order suggests that the latter might exert influence over the development of the former.  Despite the fact that the hidden order transition can only be determined for $x \leq 0.10$ \cite{Dalichaouch90}, the hidden order phase boundary extrapolates to 0~K near $x \approx 0.15$, raising the interesting possibility of the existence of a multicritical point. Whether or not the two ordered phases meet, correlations of the hidden order parameter persist into the FM phase. These have been observed in neutron scattering, as antiferromagnetic correlations and magnetic excitations associated with the hidden order phase are conspicuous for $x=0.20$, $x=0.25$, and even for $x=0.35$ \cite{Krishnamurthy08}. The effect of these correlations is uncertain, as the nature of the hidden order is still under debate and it is not known whether long-range antiferromagnetism arises at intermediate Re concentrations, as it does in the Rh case \cite{Yokoyama04}.

Our results offer insight into the origin of the unusual NFL behavior associated with the specific heat, occurring in the ordered phase \emph{below} $T_C$, a fact difficult to explain in the context of a FM quantum critical point originally identified at $x=0.30$ \cite{Bauer05}, because the NFL behavior is typically expected only outside the ordered phase \cite{Hertz76}. Based upon the present work,  the 0~K instability is identified at $x \approx 0.15$, such that NFL behavior exists only inside the FM phase, pointing to an alternative explanation. There is a notable agreement between the  values of the critical exponent $\gamma$ determined here and exponents derived from power-law fits of the bulk magnetic susceptibility of polycrystalline samples, $\chi \sim T^{-n}$ ($0.20 \leq x \leq 0.35$, $2$~K $< T < 10$~K) \cite{Bauer05,BauerPhD}. This unconventional $T$-dependence, originally attributed to NFL behavior due to a Griffiths phase \cite{Bauer05,Krishnamurthy08}, seems to actually reflect the unconventional critical scaling of the magnetic susceptibility, and strongly insinuates that the remarkable NFL behavior and critical exponents share a common origin. Both NFL behavior \cite{Affleck05} and FM order \cite{Perkins07} arise in unconventional Kondo models, implying that in URu$_{2-x}$Re$_{x}$Si$_{2}$, these seemingly exclusive effects  have a mutual origin in a novel Kondo scheme, in which Re substitution tunes the Kondo coupling between conduction electrons and uranium local moments. The proximity of the hidden order and FM phases raises the possibility that tuning of Kondo coupling may have similar influence over the stability of the inscrutable hidden order phase.

The extraordinary evolution of critical behavior in URu$_{2-x}$Re$_{x}$Si$_{2}$ underscores that the suppression of FM order is not universal, depending instead on details of competing interactions. However, similar trends in critical exponents might be found in other FM systems exhibiting a smooth suppression of $T_C$.  Their investigation will help clarify the development of criticality near continuously-tuned FM QPTs that now defies conventional understanding.

\begin{acknowledgments}
We thank B. T. Yukich for experimental assistance and D. Belitz,  S. Francoual, J. R. Jeffries, T. R. Kirkpatrick, J. Paglione, Q. Si, and S. J. Yamamoto for helpful discussions.  This work was supported by the U.S. Department of Energy (DOE) under Research Grant \# DE-FG02-04ER46105 and the National Science Foundation under Grant No. 0802478.
\end{acknowledgments}

\end{document}